# Uncertainty Principle in Control Theory, Part I: Analysis of Performance Limitations

Ji King[*]

*Abstract*—This paper investigates performance limitations and tradeoffs in the control design for linear time-invariant systems. It is shown that control specifications in time domain and in frequency domain are always mutually exclusive determined by uncertainty relations. The uncertainty principle from quantum mechanics and harmonic analysis therefore embeds itself inherently in control theory. The relations among transient specifications, system bandwidth and control energy are obtained within the framework of uncertainty principle. If the control system is provided with a large bandwidth or great control energy, then it can ensure transient specifications as good as it can be. Such a control system could be approximated by prolate spheroidal wave functions. The obtained results are also applicable to filter design due to the duality of filtering and control.

## I. Introduction

In the classical control theory, the design specifications are inherently conflicting where tradeoffs must be made to balance various control requirements. There are many different general approaches expressing the design goals or objectives for a controller design. Over past years modern sophisticated control tools have been developed to address different complex situations. These modern design methods are usually to determine a controller that minimizes a single objective or several cost functions. They may serve well to some special purposes far more than classical control techniques. But in fact these methods do not essentially settle these tradeoffs but "run the danger of obscuring insight otherwise useful for design" [1]. It is necessary and extremely important for a control system designer to give reasonable specifications and understand performance limitations in a design. The compatibility of a set of specifications, which can be simultaneously satisfied in a control system, is a feasibility problem [2]. It involves tradeoffs among competing desirable objectives, and is related to some fundamental limitations in the control theory [3].

Bode may be dated back to be the first studying feedback design limitations and tradeoffs. In his classical work [4], he stated the famous gain-phase relations and analyzed their impact upon the classical loop-shaping problem. He also derived the Bode sensitivity integral relation, which is often used to study the design limitations imposed by bandwidth constraints and open loop unstable poles. The Bode's sensitivity integral relation now has been generalized to many versions, for example, multivariable cases [5] and linear time-varying systems [6]. For details of limitations and tradeoffs in control designs, readers are referred to literatures [2], [3], [7], [8]. It is found that design limitations come from three sources: the structure of a feedback loop, realizable requirement of a control system and plant properties [8]. Under some constraints of the above, a satisfactory design, the ability to reduce the effects of load disturbances, measurement noises on the system response and the ability to track the reference signal well, will become essentially impossible. The performance limitations imposed by one or more constraints from three sources have been explicitly investigated so far [9-13], [35]. For the more new results of developments and applications of fundamental limitations，see the special issue recently published by IEEE [1].

---
[*] Ji King is with the School of Engineering, The South China Agricultural University, Guangzhou, Guangdong, 510642, China (e-mail: jiking314@gmail.com)

Within these constraints, bandwidth is the first limitation for a control system in practice. The effects of a bandwidth constraint on a control design are often considered in the frequency domain. Bode's integral relation mandated a meaningful tradeoff for frequency domain designs. It shows that it is almost always necessary to decrease open-loop gain at high frequencies to maintain sensitivity reduction against large modeling errors due to unmodeled dynamics. But sensitivity will increase in low frequencies range. Indeed, for stable open-loop systems, it is intuitively known from Bode's integral relation that the area of sensitivity reduction in a frequency range must equal the area of sensitivity increase at other frequencies. In this respect, the benefits and costs of feedback are balanced exactly. Tradeoffs must be always made between objectives. On the other hand, transient specifications in the time domain are investigated [14-18]. The widely known results are that the system's step response will exhibit undershoot due to real nonminimum phase zeros, and that the size of undershoot necessarily tends to infinity as the settling time tends to zero. It is noted that these relations among time and frequency design specifications are only obtained for low-degree systems. Their satisfactions for the general systems remain unknown. Now control techniques are still considered either in time domain or in frequency domain. And tradeoffs between time and frequency domains control requirements are not essentially involved in designs. However, a deep understanding of limitations rooted in general control systems demands a time-frequency view thus time-frequency design tools. They are expected to provide better tradeoffs and techniques from a new viewpoint.

The uncertainty principle is a meta-theorem in the time-frequency analysis. It was introduced by Gabor in 1946 from quantum mechanics to the communication [20]. Its qualitative version states that a function (signal) and its Fourier transform cannot be sharply concentrated simultaneously. There are also quantitative versions of uncertainty principle explicitly describing the relation between time concentration and frequency concentration of a function. Therefore the specifications of time domain and frequency domain can be handled with the help of uncertainty relations, as well as additional insights into the quantitative nature of tradeoffs.

A practical control system is always with bandwidth constraint. However, in this research only linear time invariant systems are considered. The signal usually refers to the impulse response of the whole control system. And the system bandwidth is relaxed to the frequency concentration of a signal. This research introduces the uncertainty principle into the control theory. The relations among time domain specifications (peak time, setting time, overshoot, and steady-state error), control energy and bandwidth (frequency concentration) are obtained based on the Slepian's version of uncertainty relation.

The remainder of this paper is organized as follows. Some results about the relations between transient specifications and bandwidth are provided in Section II. Section III presents Heisenberg's uncertainty inequality with applications to monotonic step response. In Section IV, an inequality containing time domain specifications, bandwidth and control energy is given by applying Slepian's uncertainty relation. In Section V, some other versions of uncertainty principle are presented applicable to meet different control requirements. Finally, Section VI concludes this paper.

## II. The Rise Time Bandwidth Product

In control theory, the step response is convenient to characterize process dynamics because of its simple physical interpretation. The performances on reference signal tracking of a control system can be identified from its step response. Three time domain specifications: overshoot, steady-state error and rise time of a unit step response, correspond to stability, accuracy and speed respectively imposing on a control system in the real world. The relationship among these transient performances is expected to be precise. In the filter design, the specifications for a filter are same to a control system. Hence they share the same problem in their designs. In practical communication systems, the frequency bandwidth available for data transmission is restricted, and also practical control systems are often considered to be bandlimited. Therefore, an ideal control/filter design with a desirable step response is plausible for both control and communication engineers. In the following some relations between rise time and bandwidth are presented.

The transient performances in time domain have precise relations to frequency domain specifications for simple systems. For the underdamped second order system without zeros, the rise time $t_r$ refers to the time required for a signal to change typically from 0 to 100% of its final value. $\zeta$ is the damping ratio, and $\omega_0$ is the natural frequency. Then there is $t_r\omega_0 = (\pi - \tan^{-1}(\sqrt{1-\zeta^2}/\zeta))/\sqrt{1-\zeta^2}$. A good intuition about this relation for simple systems is developed to complex systems. Consider a transfer function $H(s)$ for a stable system with $H(s) \neq 0$. The rise time is defined by the largest slope of the step response

$$t_r = \frac{H(0)}{\max_t h(t)}$$

where $h(t)$ is the impulse response of $H$, and let the bandwidth be defined as

$$\omega_b = \frac{\int_0^\infty |H(j\omega)| d\omega}{\pi H(0)}.$$

Hence

$$\max_t h(t) \leq \frac{1}{2\pi}\int_{-\infty}^{\infty}\left|e^{j\omega t}H(j\omega)\right|d\omega = \frac{1}{\pi}\int_{0}^{\infty}|H(j\omega)|d\omega$$

It now gives

$$t_r\omega_b \geq 1$$

Although this simple calculation indicates that the product of rise time and bandwidth has a lower bound, other researches shows their product is approximately constant [16], [20]. Middleton and Goodwin in [20] improved this result using the standard rule of thumb. It obtained $t_r\omega_B \approx 2.2$ where $\omega_B$ is the angular frequency with *3db* drop of magnitude known as the bandwidth commonly defined in textbooks [21]. It is argued that this approximation is true for a wide range of systems which have a "reasonable" step response. But this argument, to the authors' knowledge, hasn't been proved and improved in control field. It is clear that this product approximation cannot get a universal validation. In fact, this approximation relation is not enough for engineers to make tradeoffs among rise time, overshoot and settling time. In filtering, there are also some similar results. It is shown that the rise time $t_r$, bandwidth $\omega_r$ (defined as in [22]) and overshoot $\sigma$ of the step response of a bandlimited system are subject to the inequality $t_r\omega_r > -\ln(\sqrt{2\sigma})$. This inequality is not sharp, and its bandwidth definition is not suitable in the control practice.

### III. Heisenberg's Inequality and Optimum Monotonic Step Response

Consider the standard unity negative feedback architecture as shown in Fig.1.

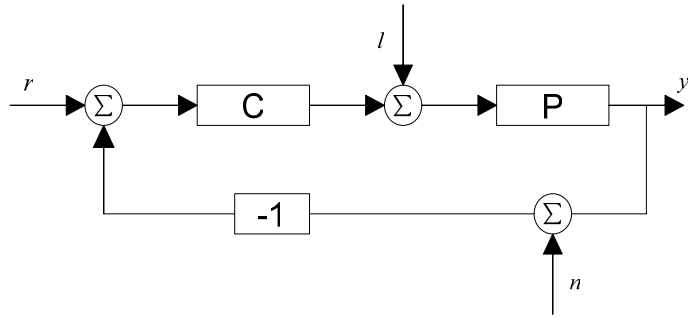

Fig. 1. Block diagram of a unity negative feedback control system

It is well known that the presence of bandwidth constraint of the plant itself and high-frequency measurement noise or load disturbances together with a bound on the desired control magnitude thus imposes a bandwidth constraint on the closed-loop system. This bandwidth cannot be supposed to be very large in order to suppress the high-frequency noise and plant uncertainty. However, the bandwidth requirement is also determined by other control objectives. The ability of reference signal tracking for a system demands a small rise time thus the large bandwidth according to their relationship presented in section II. Therefore, the performances "stability, accuracy and speed" often lead to conflicts. Let $h(t)$ denote the impulse response of a system. A precise representation of this conflicting requirement for bandwidth is that the high speed of reference signal tracking is required of sharply time concentrated $h(t)$, and the noise attenuation is required of a sharply frequency concentration $h(t)$. Then it is presented as:

How to construct a control system whose impulse response $h(t)$ is sharply time and frequency concentrated both? (1)

The answer to problem (1) is the uncertainty relation [23]. It claims that a nonzero function $h(t)$ cannot be sharply localized in time and frequency domains simultaneously. That is to say, no systems can track reference signal with very high speed and simultaneously suppress the noise very well. A precise quantitative formulation of this principle is firstly due to the celebrated Heisenberg's inequality. It can be stated as follows.

Theorem 3.1[24]. The temporal variance and the frequency variance of a signal $h \in L^2(\mathbb{R})$ satisfy

$$\sigma_t^2 \sigma_\omega^2 \geq 1/4 \qquad (2)$$

This inequality is an equality if and only if there exist $(u, \xi, a, b) \in \mathbb{R}^2 \times \mathbb{C}^2$ such that
$$h(t) = a\exp[i\xi t - b(t-u)^2]$$

Here $\sigma_t^2 = \frac{1}{\|h\|^2} \int_{-\infty}^{\infty} (t-u)^2 |h(t)|^2 dt$, $\sigma_\omega^2 = \frac{1}{2\pi \|h\|^2} \int_{-\infty}^{\infty} (\omega-\xi)^2 |\hat{h}(\omega)|^2 d\omega$ and $u$, $\xi$ are average values of $h(t)$, $\hat{h}(\omega)$ respectively. $\hat{h}(\omega)$ is the Fourier transform of $h(t)$.

The performance of tracking reference signal is usually measured by the unit step response. For a practice system, its response should be a real function. The optimal impulse response is obtained in the sense that the product of temporal variance and the frequency variance is minimal. In this case $h(t)$ are Gaussian functions and the equality in (2) is admitted. To make the impulse response normalized, let $h(t) = 2\sqrt{\frac{a}{\pi}} e^{-at^2}$. Therefore the unit step response is $u(t) = 2\sqrt{\frac{a}{\pi}} \int_0^t e^{-ax^2} dx$ with its steady-state value $u(+\infty)=1$. And its Fourier transform $\hat{h}(\omega) = 2\sqrt{\frac{a}{\pi}}\sqrt{\frac{\pi}{a}} e^{-\omega^2/4a}$ $= 2e^{-\omega^2/4a}$. The parameter $a$ is used to make a compromise between the time and frequency localizations. It is noted $u(t)$ here is a monotonic function tending to the unity. The parameter $a$ controls the speed of step response. If $a$ is large, then the step response $u(t)$ quickly tends to its steady value but $\hat{h}(\omega)$ will slowly spread in frequency range, which is bad for the high-frequency noise suppression. The unit step responses are shown in Fig.2 with different values of the parameter $a$. The speed of response is getting faster as the parameter $a$ increases. In this section the bandwidth of systems is taken as the variance of frequency response $\hat{h}$, and the rise time is the time required for the signal $h(t)$ changing from 10% to 90%. Then the product of bandwidth and rise time is a definite value. For the convenience of computations, the error function is introduced that

$$erf(t)=\frac{2}{\sqrt{\pi}}\int_0^t e^{-x^2}dx$$

It has $u(t)=erf(\sqrt{at})$. Set $u(t_1)=0.1$, $u(t_2)=0.9$. Then the rise time $t_r=t_2-t_1=1.07/\sqrt{a}$ through some calculations.

While the variance of $\hat{h}(\omega)=2e^{-\omega^2/4a}$ is $\sigma_\omega^2=2a$. The product of rise time and standard deviation is a constant. There is

$$t_r\sigma_\omega=1.52 \tag{3}$$

The equality (3) allows a tradeoff between the rise time and the bandwidth constraint. When $a=1/2$, the impulse response $h(t)$ and its Fourier transform are with the same variance. In this regard, Gaussian functions provide us an optimum system with monotonic step response. In process control a monotonically changed signal is the most common type of step response encountered. Therefore requirements of the speed of response and the ability of noise rejection can be balanced by the accurate formula.

If the setting time is defined as the duration that is required the signal being within the range of 3% of its steady value, then it has $t_s=1.53/\sqrt{a}$. When the settling time is adopted as the measure of speed, there is a new relation that $t_s\sigma_\omega=2.17$.

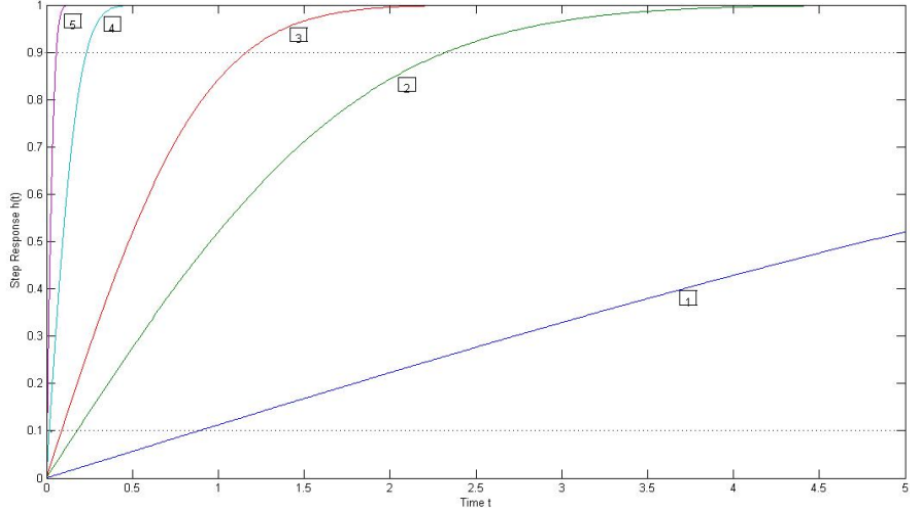

Fig. 2. Unit step response $u(t)$ indexed by 1-5 of the system with impulse response $h(t)=2\sqrt{a/\pi}\exp(-at^2)$ for $\sqrt{a}=0.1, 0.5, 1, 5, 20$. As $a$ increases, the graph of $u(t)$ tends sharply to the unity toward the $t$-axis.

IV. Prolate spheroidal wave functions

The variance measure of signal concentration and its induced optimal step response have applications in some types of control system designs. However, it can fail in many other situations. The crawling dynamic of monotonic step response is regarded to be bad in some control systems. On the other hand when the plant to be controlled is bandlimited, the whole system is also with a finite bandwidth whatever the controller is selected. For example, a controller design for the communication channel, subject to the Bode's integral formula, is such a plant under the finite bandwidth constraint [25]. Moreover the system is also required to track the reference signal in a very finite elapsed time. Hence the time and frequency concentrations in problem (1) are needed to be redefined with the length of time and frequency segments occupied in their respective ranges. It is therefore expected to construct a control system whose impulse response is timelimited and bandlimited both. It is known that is impossible except the trivial always zero signal, which makes no sense to designers.. But signals that are synthesized from a finite frequency band and lasting for a finite length of time indeed always exist in the real world including control systems [26].

One needs a new comprehension of these time- and band-limited signals. Fortunately this problem had been much studied in communication [27-29], and even can be traced back to works in mathematics many years ago [30]. Instead of variance, a natural and meaningful measure of concentration of signal is the fraction of the signal's energy that lies in a given time slot

$$\alpha^2(T) = \frac{\int_{-T/2}^{T/2} h^2(t)dt}{\|h\|^2}. \tag{4}$$

Set $P_T$ and $Q_W$ operators truncating $h$ and $\hat{h}$ respectively that $P_T h = h\chi_T$, $(Q_W h)^{\wedge} = \hat{h}\chi_W$. Here $\chi_T$ and $\chi_W$ are indicator functions of sets $(-T/2, T/2)$ and $(-W, W)$. Similarly in frequency range there is

$$\beta^2(W) = \frac{\int_{-W}^{W} |\hat{h}(\omega)|^2 d\omega}{\|\hat{h}(\omega)\|^2}. \tag{5}$$

It has $\|P_T h\|_2 = \alpha$ and $\|Q_W h\|_2 = \beta$. That a signal bandlimited within $(-W, W)$ and time concentrating in $(-T/2, T/2)$ with its largest value $\alpha^2(T)$ is subject to the following integral equation

$$\frac{1}{2\pi}\int_{-W}^{W}\frac{\sin(\xi-\xi')T/2}{(\xi-\xi')/2}\hat{h}(\xi')d\xi' = \alpha^2(T)\hat{h}(\xi), \quad |\xi|\leq W. \tag{6}$$

Make the substitutions $\xi = W\omega$, $\xi' = W\omega'$, $\hat{h}(W\omega) = \psi(\omega)$. The equation (6) reduces to

$$\int_{-1}^{1}\frac{\sin c(\omega-\omega')}{\pi(\omega-\omega')}\psi(\omega')d\omega' = \lambda\psi(\omega), \quad |\omega|\leq 1 \tag{7}$$

in which $c = WT/2$, $\lambda = \alpha^2(T)$.

This integral equation (7) contains a countable number of eigenvalues. Set $1 > \lambda_0 > \lambda_1 > \ldots$ and it tends to zero as $n \to \infty$. The solutions of (7) corresponding to their eigenvalues are $\psi_0(\omega), \psi_1(\omega), \ldots$ . These eigenfunctions are called prolate spheroidal wave functions (PSWFs). $\{\psi_n\}$ are orthogonal and complete in $L^2(-1,1)$. From the duality of time and frequency in the Fourier transform, PSWFs are also functions occupying the maximal signal energy fraction over a frequency band given these signals are timelimited. Actually it has

$$\frac{1}{2\pi}\int_{-1}^{1}\psi_n(\omega)e^{i\omega t}d\omega = \alpha_n\psi_n(t/c), t\in \mathbb{R}.$$

Here $\alpha_n$ is a constant associated with the order $n$ but independent of $t$. More properties of PSWFs can be found in [30]. For a general signal that is neither timelimited nor bandlimited, then given a pair $(\alpha, \beta)$ the concentration in time and frequency domains, whether there is the corresponding function $h(t)$ determined by prolate spheroidal wave functions? There is the following theorem [23].

Theorem 4.1. Suppose $0 \leq \alpha, \beta \leq 1$ and $(\alpha, \beta) \neq (1, 0)$ or $(0, 1)$. There is a function $h \in L^2(\mathbb{R})$ with $\|P_T h\|_2 = \alpha$, and $\|Q_W h\|_2 = \beta$ if and only if

$$\arccos\alpha + \arccos\beta \geq \arccos\sqrt{\lambda_0} = \arccos\|P_T Q_W\|_2. \tag{8}$$

This theorem describes the tradeoff between signal energy concentrations in time and frequency ranges simultaneously for a general finite energy signal, and it is regarded as another quantitative statement of uncertainty principle. It is noted that in its full statement this theorem is complicated and $h(t)$ is required to has unit energy, which can be found in [28]. In fact, this condition can be removed without loss of generality. Concerning the signal energy ratio $\alpha$ and $\beta$, there is another theorem [31].

Theorem 4.2. Suppose $(\alpha, \beta)$ is an admissible pair, i.e., suppose that $\alpha = \|P_T h\|_2$ and $\beta = \|Q_W h\|_2$ for an actual function $h$ of unit energy. Then any pair $(\alpha_1, \beta_1)$ with $0 < \alpha_1 \leq \alpha$ and $0 < \beta_1 \leq \beta$ is also admissible.

In the following, the inequality (8) is used to obtain the relation among bandwidth and transient specifications. The Cauchy–Schwarz inequality is introduced firstly.

$$(\int_0^T fg\, dx)^2 \leq (\int_o^T f^2 dx)(\int_0^T g^2 dx) \tag{9}$$

Set $f = h$, $g = 1$, and it has

$$(\int_0^T h(t)dt)^2 \leq T\int_0^T h^2(t)dt \tag{10}$$

When $h(t)$ is a constant ($0 \leq t \leq T$), the equality in (10) is obtainable.

Denote the energy $E = \|h\|_2^2 = \int_{-\infty}^{\infty}|h(t)|^2 dt$ and the step response at time $T$ is $u(T) = \int_0^T h(t)dt = 1 + \delta(T)$, where $h(t)$ is normalized so that $u(+\infty) = 1$ and $\delta(T)$ is the deviation to the steady-state value at $t = T$. The ratio of time signal energy concentrated on $(0, T)$ to the whole energy is denoted by $\alpha^2(T)$. Then there is $\int_0^T h^2(t)dt = E\alpha^2(T)$. Substituting these notations into (10), it becomes

$$\frac{|1+\delta(T)|}{\sqrt{ET}} \leq \alpha(T) \leq \sqrt{\lambda_0} \tag{11}$$

To simplify the discussion, $u(t)$ is supposed to be positive as $t > 0$. From (11) a new inequality is obtained

$$\arccos\frac{1+\delta(T)}{\sqrt{ET}} \geq \arccos\alpha(T) \qquad (12)$$

Combining (8) and (12), there is

$$\arccos\frac{1+\delta(T)}{\sqrt{ET}} + \arccos\beta \geq \arccos\sqrt{\lambda_o} \qquad (13)$$

Equality in (8) is attained for the function $h(t)=p\psi_0+qP_T\psi_0$, with

$$p = \sqrt{(1-\alpha^2)/(1-\lambda_0)}$$

and

$$q = \alpha/\sqrt{\lambda_0} - p.$$

But conditions for the equalities in (8) and (10) are different. Therefore the equality of (13) is not admitted. It becomes

$$\arccos\frac{1+\delta(T)}{\sqrt{ET}} + \arccos\beta > \arccos\sqrt{\lambda_o} \qquad (14)$$

Its simplified version is

$$\frac{1+\delta(T)}{\sqrt{ET}} < \sqrt{\lambda_o} < 1 \qquad (15)$$

Now it has established the relation with the help of (14) among time $T$, signal (control) energy $E$, step response deviation $\delta(T)$ and the bandwidth constraint measured by $\beta$. For control systems, the important performances "stability, accuracy, speed" are always required. The stability can be intuitively measured by the overshoot since an unstable system will lead to a large, even infinite deviation $\delta(T)$ to the steady-state value. The step response can be normalized by choosing a suitable proportion so that its steady state error is null and the accuracy is then promised. But the speed of step response is not readily measured. By the usual definitions the rise time or setting time is difficult to be evaluated for the general linear time-invariant systems. However, it is noted the peak time $t_p$ is always subject to $t_r<t_p<t_s$. Obviously all the time domain specifications ($t_r, t_p, t_s, \delta$) cannot be encapsulated into a relation. It is reasonable to take the intermediate quantity $t_p$ as the measure on speed of the system response. If set $T=t_p$ in (14), then three important time domain specifications and bandwidth are enclosed in an inequality. This inequality provides us a method validating reasonable specifications. And with this inequality tradeoffs between different requirements are freely made to achieve suitable performances. The fundamental limitation in a LTI control system is hence imposed by (14).

In the filtering field the frequency-selective filter design needs specifications in the frequency domain which are the counterparts of step response of control systems in time domain. So this inequality here is also applicable to the filter design.

In the following the inequality (14) will be discussed. These specifications and the implications of (14) will be refined.

*A. On the Largest Eigenvalue $\lambda_0$*

In the inequality (14), $\lambda_0$ is an important parameter depending on the bandwidth $W$ and time segment $T$. It is the largest eigenvalue of the compact operator associated with integral equation (7). When $T$ or $W$ is infinite, the kernel function in (6) and (7) becomes a reproducing kernel therefore $\lambda_0$ equals to the unity. This eigenvalue is the definitive value that the greatest energy concentration can occupy. For practical considerations, this eigenvalue and its relation to the product $WT$ should be quantitatively analyzed. There are some results about the asymptotic behavior of eigenvalues of PSWFs [32], [33].

Theorem 4.3 [32]. Let $\lambda_0 > \lambda_1 > \lambda_2 \ldots$ be the eigenvalues of the integral equation (7). Then

$$1 - \lambda_n \sim 4\pi^{1/2} 8^n (n!)^{-1} c^{n+1/2} e^{-2c}, \text{ as } c \to \infty \tag{16}$$

Set $n=0$ substituting into formula (16). There is an approximation

$$\lambda_0 \sim 1 - 4\sqrt{\pi c}\, e^{-2c} \tag{17}$$

The plot of this approximation is shown in Fig. 3. From the chart it is seen that the formula (17) can well approximate to the eigenvalue $\lambda_0$ when $c$ is large.

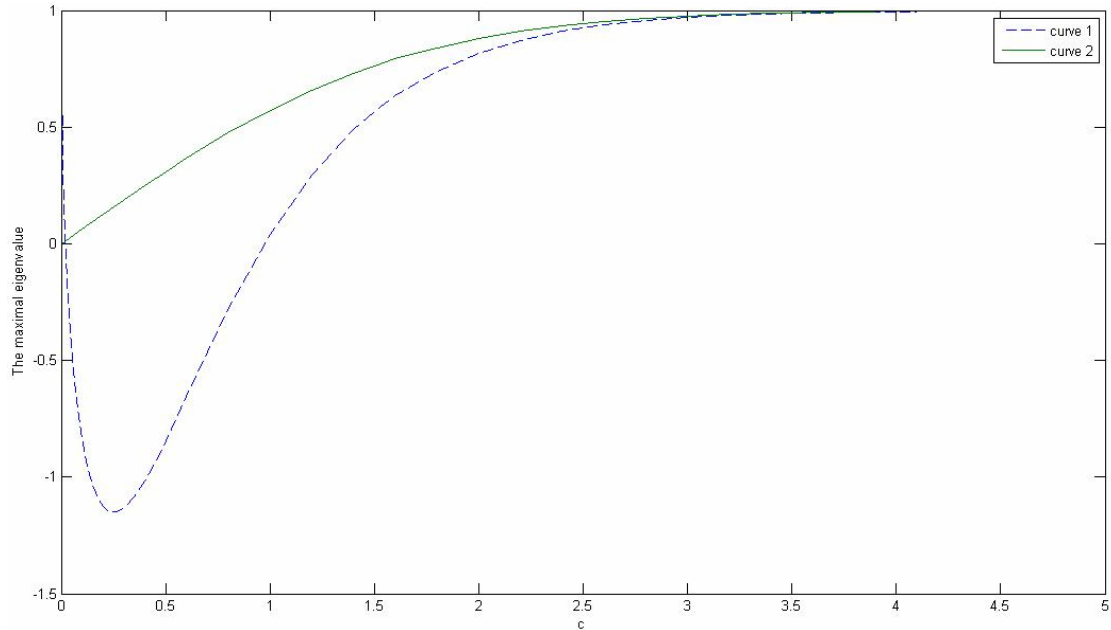

Fig. 3. The largest eigenvalue $\lambda_0$ corresponding to PSWF in terms of $c=WT/2$. Curve 1 depicts the accurate eigenvalue and Curve 2 is its approximation.

*B. On the Bandwidth and the Energy*

An ideal step response should be quickly decaying (small setting time) but simultaneously with high rising speed and very little overshoot. If these specifications are subject to the inequality (14), then there is a step response thus a control system satisfied with the given specifications. The signal can be approximated by the complete function set $\{\psi_0, \psi_1, \ldots\}$ only if its specifications are subject to (14). In this sense, it says (14) is sharp. But this "sharpness" is indeed obtainable at the cost of more bandwidth or more control energy of systems. Here the bandwidth refers to $\beta$, the energy concentration in frequency domain, and signal (control) energy $E$ of $h(t)$ is contributed by all components of a system, controller $C$, plant $P$ and feedback $H$. Then the problem is raised that whether ideal transient performances are achievable when the bandwidth and energy are given beforehand. Let's return to the inequality (14) and (15) again. It has been known the equality in (8) is achievable when $h(t)=\psi_0$. This zeroth order PSWFs with different parameters $c$ are shown in Fig. 4 and its refinement is shown in Fig. 5. The oscillation of PSWFs will make step response of the system (its impulse response function $h(t)=\psi_0$) also change around its steady-state value with very long time. Therefore PSWFs are far from an ideal impulse response.

For an ideal step response, it should be $1+\delta(T) \sim 1$, $T$ is small, hence $(1+\delta(T))/\sqrt{ET}$ is close to 1 and $\arccos\dfrac{1+\delta(T)}{\sqrt{ET}}$ is very small. While $\lambda_0$ depends on $WT$, it decreases when $T$ becomes small. So $T$ cannot be arbitrarily small given $\beta$, otherwise (14) and (15) might be violated. However, if bandwidth and energy are provided large enough, then it has $\arccos\beta \to 0$, $\lambda_0 \to 1$. It leaves designers room to adjust these specifications. The arbitrary good transient performances can be acquired through enlarging bandwidth or energy in theory.

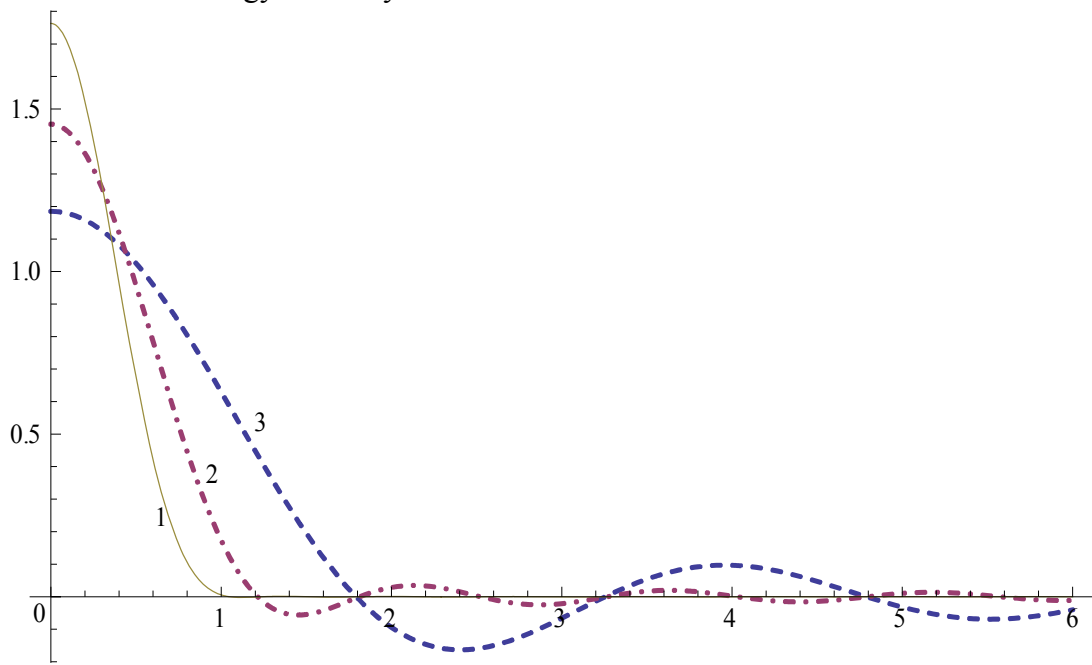

Fig. 4. The zeroth prolate spheroidal wave function $\psi_0$, the curves 1, 2, 3 for $c=8, 4, 2$. The magnitude of change in curve 1 for $t>1$ is very small and cannot be shown in the scales for curve 2, 3.

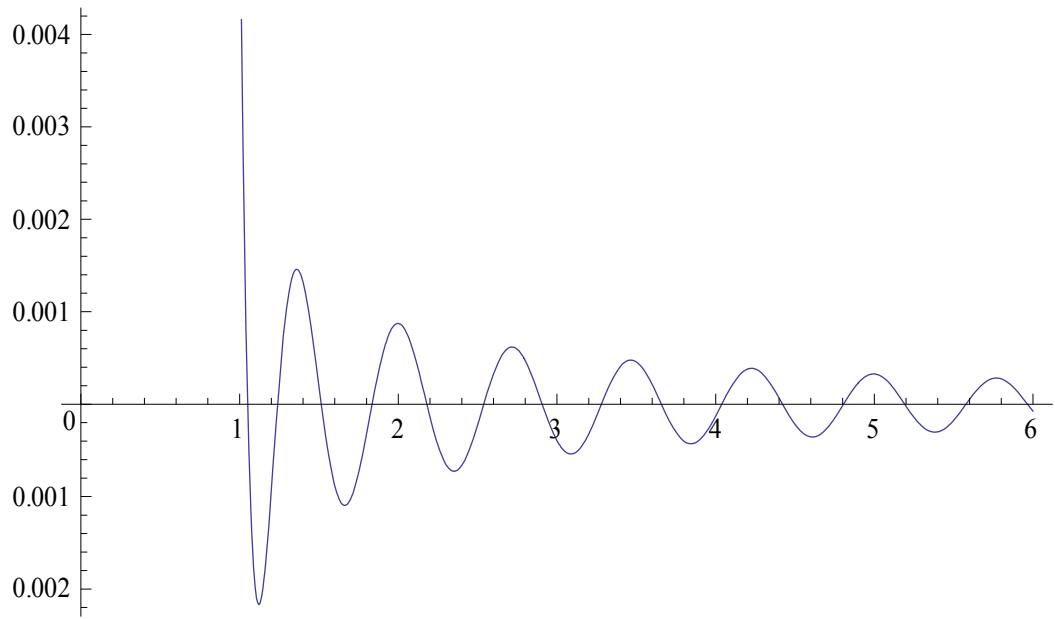

Fig. 5. The oscillation of $\psi_0$ for $c=8$, $t>1$.

But the method injecting more signal energy into a system will cause side effects than enlarging system bandwidth. By the ideal impulse response mode as expected in the above, it should be in form of the Mexican hat, which is depicted in Fig. 6. The zero crossing point is the peak time and the integration of this signal from origin to the zero crossing point corresponds to the overshoot in the step response. If a very small peak time $t_p$ and very little overshoot $\delta(t_p)$ are commanded, then the signal energy within $(0, t_p)$ should also be small as the magnitude of $h(t)$ is restricted for the reason of stability. However, by the large energy it requires the large energy of $h(t)$ for $t>t_p$ to compensate the small energy for $t<t_p$. Therefore it needs much time for $h(t)$ converging to zero. The setting time is then not so small. And when the small setting time is requested, $h(t)$ will oscillate to ensure the enough energy and unity steady-state value also shown in Fig. 6. This phenomenon exactly agrees with one's experience in control designs. For example, a high proportional gain results in a faster response, and the large derivative value is used to reduce the magnitude of the overshoot in a proportional-derivative (PD) controller. Hence a PD controller may response quickly and has a small overshoot simultaneously. But the proportional and derivative gains add great energy to the control system. Excessively large PD values will lead to oscillation even instability, which is consistent with the above discussions.

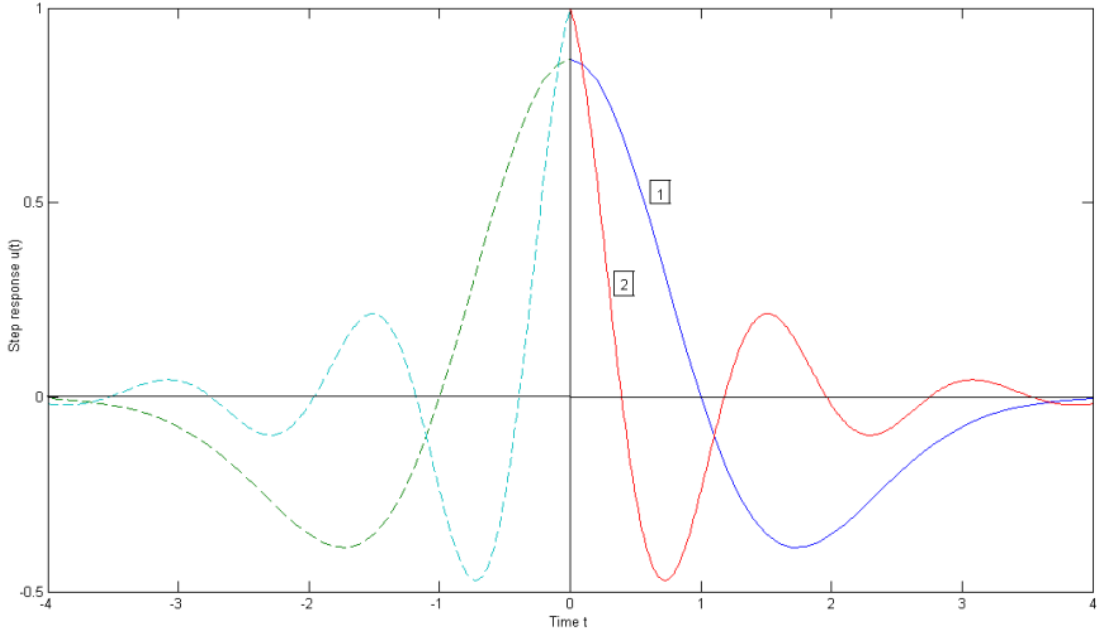

Fig. 6. Curve 1 is an ideal impulse response in the Mexican hat form. Curve 2 must oscillate to preserve the enough energy but with small peak time and very little overshoot.

When the plant *P* is bandlimited, the only way to get a good transient performance is adding energy. But in other cases, for example, the bandwidth is defined to be the angular frequency with *3db* drop of magnitude, injecting energy into controller *C* can also enlarge the bandwidth. However, bandwidth, which is intrinsic characteristic in control systems and communications, is essentially different from energy.

*C. Transient Analysis*

An ideal low-pass filter is a simple bandlimited system. For such a system, its rise time is inversely proportional to the bandwidth. And the overshoot is exact the Gibbs phenomenon. It is well known that an ideal low-pass filter is not physically realizable. Therefore a new function $\hat{h}(\omega)$ is applied instead of the rectangle window to suppress Gibbs oscillations, for example, the step response of a triangle window doesn't have an overshoot.

It will be shown below the difference between peak time $t_p$ and rise time $t_r$ is very slight for a response in the Mexican hat model. Here the rise time refers to the time required for a signal to change from the origin to 100% of the step height. As $u(t_r)=\int_0^{t_r} h(t)dt=1$, it has $\bar{h}_r t_r =1$. And $u(t_p)= \int_0^{t_p} h(t)dt = \int_0^{t_r} h(t)dt + \int_{t_r}^{t_p} h(t)dt =1+\delta(t_p)$, it has $\bar{h}_p t_p = \bar{h}_r t_r + \bar{h}(t_p - t_r)$. $\bar{h}_r, \bar{h}_p$ and $\bar{h}$ are the averages of $h(t)$ in $(0, t_r)$, $(0, t_p)$ and $(t_r, t_p)$ respectively. But from the convexity of ideal impulse response, it is readily known $\bar{h} > \bar{h}_p > \bar{h}_r$. There is $t_p - t_r = \delta(t_p)/\bar{h} < \delta(t_p)/\bar{h}_p = \delta(t_p)t_p/(1+\delta(t_p)) \approx \delta(t_p)t_p$. Under the popular specification $\delta(t_p)=5\%$, this difference is negligible for common control systems.

For a normalized impulse response with the Mexican hat mode, there is

$$\left|\int_{t_p}^{\infty} h(t)dt\right| \geq \int_{t_p}^{\infty} |h(t)|^2 dt \quad \text{as } |h(t)| < 1, (t > t_p). \tag{18}$$

Then it has

$$1 + \left|\int_{t_p}^{\infty} h(t)dt\right| = \int_0^{\infty} h(t)dt - \int_{t_p}^{\infty} h(t)dt = \int_0^{t_p} h(t)dt \geq 1 + \int_{t_p}^{\infty} h^2(t)dt = 1 + E(1-\alpha^2(t_p))$$

There is

$$\left|\int_{t_p}^{\infty} h(t)dt\right| \geq E(1-\alpha^2(t_p)) \tag{19}$$

The overshoot is

$$\int_0^{t_p} h(t)dt - 1 = \left|\int_{t_p}^{\infty} h(t)dt\right| \geq E(1-\alpha^2(t_p)) \tag{20}$$

As $\alpha^2(t_p) \leq \lambda_0(t_p)$, then $\delta(t_p) \geq E(1-\lambda_0(t_p))$ $(t \geq t_p)$. The minimal deviation (error) is then obtained.

*D. On One-sided Transform Effect*

The time concentration of impulse response is considered so far within the positive axis because of the physical significance of a real control system. But note that the time concentration and frequency sharpness are both defined in symmetric intervals as in (4) and (5). This symmetry leads to beautiful integral equations (6), (7) and their elegant real PSWFs solutions. If the symmetry is broken, then the optimal signal that has the largest energy fraction in time domain locating within a finite frequency range becomes a complex function, which isn't suitable for practical applications. However, the real system is casual and its response is only excited at $t \geq 0$. And real frequencies also only exist in the positive range. One wonders whether the inequality (14) and deductions would be violated. It is shown that the PSWFs from (6) and (7) are symmetric that

$$\alpha^2(T) = \int_{-T/2}^{T/2} |\psi_i(t)|^2 dt / \int_{-\infty}^{\infty} |\psi_i(t)|^2 dt = \int_0^{T/2} |\psi_i(t)|^2 dt / \int_0^{\infty} |\psi_i(t)|^2 dt.$$

For a given frequency range $(0, W)$ and time interval $(0, T/2)$, the signal can be extended symmetrically to its counterparts $(-W, 0)$ and $(-T/2, 0)$. Therefore the signal is again defined in the whole time and frequency domains. It is clear the PSWFs behave at $t \geq 0$ as in the whole axis. That is to say, the inequality (14) and deductions are also true when the positive parts of PSWFs are employed.

V. More Uncertainty Relations

Uncertainty principle is a general statement about universal conjugate variables in harmonic analysis and quantum mechanics. It has been found playing in many areas where a great variety of uncertainty relations are presented. In additional to Heisenberg's inequality and Slepian's PSWFs there are some other relations. In this section some relations that are suitable for some special purposes in control system design will be introduced. For transient analysis and performance tradeoffs these uncertainty relations have their own benefits.

In [28] for a constraint, there is another "interesting phenomenon" for $\alpha$ and $\beta$.

**Theorem 5.1.** If $\alpha^2+\beta^2\leq 1$, then $\arccos\alpha + \arccos\beta \geq \pi/2$.

Since $\arccos\sqrt{\lambda_0} \leq 1 < \pi/2$ for any $c=WT/2$, the above inequality will exceed $\arccos\sqrt{\lambda_0}$. As a consequence, this theorem can be quickly transformed to relations among transient specifications and bandwidth. In conformity with notations in Section IV, there is the following theorem from (11) and (12).

**Theorem 5.2.** If $\dfrac{(1+\delta(T))^2}{ET}+\beta^2\leq 1$, then $\arccos\dfrac{1+\delta(T)}{\sqrt{ET}}+\arccos\beta \geq \pi/2$.

In the previous section it is known that the inequality (14) is sharp and PSWFs are also applicable to the casual control systems. However, the inverse cosine functions in (14) make this relation complicated and performance analysis on arbitrary intervals is still unavailable. There is Chalk's uncertainty relation [28], [34] providing a relative concise inequality to address asymmetric time or frequency segments. Also it doesn't definitely need signals to be $L^2$ integrable. This relation was also considered to be firstly made by L.A. MacColl around 1940 in an unpublished communication [28]. The new energy portions concentrating in time and frequency domains are that

$$\frac{\int_{t_0}^{t_0+T}|h(t)|^2\,dt}{\int_{-\infty}^{\infty}|h(t)|^2\,dt}=\alpha_1 \tag{21}$$

and

$$\frac{\int_{w_0}^{w_0+W}|\hat{h}(\omega)|\,dt}{\int_{-\infty}^{\infty}|\hat{h}(\omega)|\,d\omega}=\beta_1 \tag{22}$$

There is $WT>2\pi\alpha_1\beta_1$.

For $\int_{t_0}^{t_0+T}|h(t)|^2\,dt=\alpha_1 E\geq\dfrac{(1+\delta(T))^2}{T}$, it has

$$WT>2\pi\alpha_1\beta_1\geq 2\pi\frac{(1+\delta(T))^2}{ET}\beta_1 \tag{23}$$

To write (23) in a concise form, it becomes

$$EW(\frac{T}{1+\delta(T)})^2>2\pi\beta_1 \tag{24}$$

On the duality of time and frequency domains, also set

$$\frac{\int_{t_0}^{t_0+T}|h(t)|dt}{\int_{-\infty}^{\infty}|h(t)|dt}=\alpha_1' \tag{25}$$

and

$$\frac{\int_{w_0}^{w_0+W}|\hat{h}(\omega)|^2 dt}{\int_{-\infty}^{\infty}|\hat{h}(\omega)|^2 d\omega}=\beta_1' \tag{26}$$

It has $WT > 2\pi\alpha_1'\beta_1'$ (27)

Denote the $L^1$ norm $\int_{-\infty}^{\infty}|h(t)|dt = E_1$. If $T<t_p$ and $h(t)$ is positive, then

$$1+\delta(T)=\int_0^T|h(t)|dt=\alpha_1'E_1 \tag{28}$$

So

$$\alpha_1'=\frac{1+\delta(T)}{E_1} \tag{29}$$

Substituting (29) into (27), it has

$$E_1W(\frac{T}{1+\delta(T)})>2\pi\beta_1' \tag{30}$$

The inequalities (24) and (30) again indicate the importance of bandwidth $W$ and energy $E_1$ or $E$. One can then apply (24) and (30) to examine readily the compatibility of control specifications.

The uncertainty relations mentioned so far called global uncertainty relations seem don't preclude $h$ and $\hat{h}$ from being concentrated in a small neighborhood of two or more widely separated points. But unfortunately local uncertainty inequalities tell us it is impossible for this phenomenon [23]. Also $h$ and its Fourier transform $\hat{h}$ cannot very sharply locate in a small neighborhood of points both.

Theorem 5.3. (i). If $0<a<1/2$, there is a constant $K_a$ such that for all $h \in L^2(\mathbb{R})$ and all measurable $W \subset \mathbb{R}$,

$$\int_W |\hat{h}|^2 d\omega \leq K_a |W|^{2a} \left\| |t|^a h \right\|_2^2.$$

(ii). If $a>1/2$, there is a constant $K_a$ such that for all $h \in L^2(\mathbb{R})$ and all measurable $W \subset \mathbb{R}$,

$$\int_W |\hat{h}|^2 d\omega \leq K_a |W| \|h\|_2^{2-1/a} \left\| |t|^a h \right\|_2^{1/a}.$$

This theorem is also satisfied when $h$ and $\hat{h}$ are interchanged. Although it is a quantitative uncertainty relation, it remains so "qualitative" for the purpose of specification examinations. However, the two parts (i) and (ii) of the theorem both indicate the maximal energy of $\hat{h}$ (or $h$) in an interval is proportional to a certain power of length of this interval. Now set $a=1$, and extend $\hat{h}$ to be an even function. It has

$$\int_T |h|^2 dt \leq K_a |T| \|\hat{h}\|_2 \|\omega|\hat{h}|\|_2 \tag{31}$$

for a constant $K_a$ and an interval $(0, T)$ denoted by $T$.

There is $\|\hat{h}\|_2 \|\omega|\hat{h}|\|_2 = \sigma_\omega \|\hat{h}\|_2^2$, where $\sigma_\omega$ is defined in the Theorem 3.1 and expectation of $\hat{h}$ is zero for it is an even function. The inequality (31) becomes

$$\int_T |h|^2 dt \leq K_a' \sigma_\omega ET, \tag{32}$$

where $K_a' = 2\pi K_a$. Combining (10) and (32), it has

$$\frac{1+\delta(T)}{T} \leq \sqrt{K_a' \sigma_\omega E} \tag{33}$$

This inequality can be considered as a qualitative description of transient specifications and the energy.

## VI. Conclusion and Discussion

This paper has obtained quantitative results to examine the consistence of different control requirements including transient performances and bandwidth tradeoffs. It is found the uncertainty principle borrowed from quantum mechanics and harmonic analysis bridges the time and frequency domains and reaches to a compromise among specifications in the two domains. The concrete relations of uncertainty principles give various interpretations about the performance limitations and tradeoffs in control designs.

Based on these relations, the optimum monotonic step response is achieved through Heisenberg's inequality. By introducing Slepian's uncertainty principle, specifications on the performances "stability, accuracy and speed" of systems in time domain, the control energy and bandwidth constraint are encapsulated into a sharp inequality. Only system specifications which are satisfied with this inequality can be realized in a control design. And this system realization can be approximately obtained by PSWFs. The system performance is also improved by enlarging bandwidth and injecting more control energy into the system as suggested by the inequality.

In addition, more uncertainty relations are discussed in occasions of control designs for convenience. It is shown that asymmetrical intervals can be well dealt with applying Chalk's uncertainty relation. And the local uncertainty inequality provides us a qualitative examination of time and frequency domain specifications.

As seen in the previous discussions, the conflicting requirements of a system are recognized by the investigation the inequality (14) in this system. These inequalities are the best available results under their respective conditions for linear time-invariant systems. In a controller design the performances can be degenerate because of plant properties, perturbations from the environment and other constraints. Further work will be dedicated to applying results of this paper to quantitative controller designs. Possible technical details would be the approximation of controller to get an ideal response, noise attenuation, addressing bad plant properties.